# Delay of Leidenfrost point during drop impact of surfactant solutions

Gudlavalleti V V S Vara Prasad [a], Purbarun Dhar [b,*,1] and Devranjan Samanta [a,*,2]

[a] Department of Mechanical Engineering, Indian Institute of Technology Ropar, Punjab–140001, India

[b] Hydrodynamics and Thermal Multiphysics Lab (HTML), Department of Mechanical Engineering, Indian Institute of Technology Kharagpur, West Bengal–721302, India

*Corresponding authors:

[1]E-mail: purbarun@mech.iitkgp.ac.in; purbarun.iit@gmail.com

[1]Tel: +91-3222-28-2938

[2]E-mail: devranjan.samanta@iitrpr.ac.in

[2]Tel: +91-1881-24-2109

## Abstract

In this article, a novel method of increasing the dynamic Leidenfrost temperature $T_{DL}$ is proposed through the addition of both anionic (SDS) and cationic (CTAB) surfactants to water droplets. We focus on understanding the hydrodynamics and thermal aspects of droplet impact Leidenfrost behaviour of surfactant solutions, and aim to delay the onset of the Leidenfrost regime. The effects of Weber number (*We*), Ohnesorge number (*Oh*) and surfactant concentration on dynamic Leidenfrost temperature ($T_{DL}$) were experimentally studied in details, covering a wide gamut of governing parameters. At a fixed impact velocity, $T_{DL}$ increased with the increase of surfactant concentration. $T_{DL}$ decreased with increase of impact velocity for all solutions of surfactant droplets at a fixed surfactant concentration. We proposed a scaling relationship for $T_{DL}$ in terms of *We* and *Oh*. At temperatures (~ 400°C) considerably higher than $T_{DL}$, droplets exhibit trampoline like dynamics or central jet formation, associated with fragmentation, depending upon the impact velocity. Finally, a regime map of the different boiling regimes such as transition boiling, Leidenfrost effect, trampolining and explosive behaviour is presented as function of impact *We* and substrate temperature ($T_s$). The findings may hold strong implications in thermal management systems operating at high temperatures.

# 1. Introduction

Liquid-vapour phase change is ubiquitous in several natural phenomena and man-made utilities. Under its wide aegis, the Leidenfrost effect is an atypical phenomenon where the liquid may levitate above its own vapour cushion at temperatures significantly higher than its boiling point [1]; often observed in day-to-day life when water droplets come in contact with superheated kitchen-ware. The corresponding temperature at which the onset of the effect takes place is known as the Leidenfrost point (or temperature). At this juncture, the heat transfer rate between the fluid and the heated surface drops significantly as the vapour cushion formed beneath the fluid droplet is a poor thermal conductor. Impact of liquid droplets on such superheated surfaces are of prime relevance to many industrial processes such as spray quenching of alloys, fire extinguishment by sprinkler systems [2] additive manufacturing and coatings [3], and turbine blade cooling [4]. The reduction in heat transfer can be detrimental in several such cases, for example in the thermal management of nuclear power plants [3] or metallurgical treatments. The droplet Leidenfrost effect also has great potential in drag reduction [5] as well as nanoscale manufacturing processes [6].

In recent times, there have been significant strides in the field of droplet Leidenfrost dynamics. Quere and Clanet et al have made seminal contributions in this field [1, 7-10]. Biance et al studied the shape of Leidenfrost droplets and its dependence on the dynamics of the vapour layer beneath it [7]. They also characterized the lifetime of the Leidenfrost droplets. Self propulsion of Leidenfrost drops [8] or solids (those which directly sublimate over heated substrates [9]) can be controlled with specially designed ratcheted sufaces. Directed motion in Leidenfrost drops can be useful for efficient heat transfer [8-9] and targetted hot-spot cooling. Dupeux et al [10] incorporated crenulations on the surface to increase friction and control the mobility of Leidenfrost droplets. Bouillant et al [11] performed particle image velocimetry (PIV) within the Leidenfrost droplets (at 300 $^o$C) and established that the mobility is dependent on the droplet sizes. They showed that if the droplet diameters are lesser than the capillary length scale, only one vortex exists within the drop. On the contrary, larger diameter droplets showed the presence of two internal vortices. Shirota et al [12] explored the relevant length and time scales vital for the transition from the fully wetted to the fully levitated (i.e. Leidenfrost) state of droplets on an isothermal hot substrate using total internal reflection (TIR) imaging technique.

Tran et al [13] proposed a new scaling relationship for dimensionless maximum spreading ($\gamma$) of impacting droplets on the heated surfaces in both gentle and spray film boiling regimes with Weber number ($\gamma \sim We^{0.4}$). They showed that the spreading dynamics for heated conditions is higher compared to that for ambient conditions ($\gamma \sim We^{0.25}$). Khavari et al. [14] experimentally investigated the impact dynamics of droplets on sufficiently heated surfaces and based on the fingering patterns of post impact droplet, classified the boiling into four regimes, viz. spreading (negligible heating effect), bubbly boiling (activated nucleate boiling at solid-liquid interface), fingering boiling, and Leidenfrost regime, respectively. Villegas et al [15] performed direct numerical simulations of the Leidenfrost state and studied the temporal evolution of drop shapes and vapour layer thickness using the level set and ghost fluid methods. Qiao et al [16] used lattice Boltzmann techniques to simulate the Leidenfrost dynamics of droplets over heated liquid pools. They explored the effect of various non-dimensional parameters like vapour Stefan number, Bond number ($Bo$), Ohnesorge number ($Oh$) and releasing height, and pool depth on the Leidenfrost dynamics.

In addition to the research aimed at understanding of fundamental aspects of droplet Leidenfrost effect, there have been efforts to modulate the dynamic Leidenfrost temperature ($T_{DL}$) of fluids via addition of non-Newtonian additives [17–20]. It has been shown by Dhar et al [21] that when long chain polymers are dissolved in Newtonian fluids, the resultant elastic fluid droplets exhibit rebound suppression on superhydrophobic surfaces. This rebound suppression behaviour has been exploited in the case of Leidenfrost droplets also, with the aim to arrest rebound and thereby delay the reduction of heat transfer calibre of the droplets. Secondary atomization of droplets and increase of $T_{DL}$ was observed by Bertola and his co-workers [17–19] with polymer additives in the droplet fluid. Dhar et al [20] highlighted the role of Weber number and polymer concentration on the Leidenfrost point (LFP). They showed that the long-lasting filaments attached to the substrate are responsible for the increase of $T_{DL}$. The elastic effects of polymer chains in suppressing the droplet rebound at heated [20] and ambient conditions [21] were explored by Dhar et al. and was shown to be potential method to delay the Leidenfrost effect in droplets.

Another route to morph the $T_{DL}$ is through the addition of surfactants, which can alter the surface tension and the wetting characteristics of the fluid. Qiao and Chandra [22] performed one of the earliest studies on the effect of common surfactants (sodium dodecyl sulphate, SDS) on drop impact behaviour at high surface temperatures. Addition of surfactants enhanced the nucleation of vapour bubbles, reduced bubble coalescence and promoted foaming in the liquid. Increase in surfactant concentration led to decrease in the $T_{DL}$. The surfactants could not affect the evaporation time of the droplets in the film boiling regime, but spray cooling regime was enhanced by addition of surfactants [23]. In contrary to Qiao and Chandra's observations [22], Chen et al. [24] reported the increase of $T_{DL}$ by addition of high alcohol surfactants (HAS). They established an empirical correlation between $T_{DL}$ and the maximum spreading factor. Zhang et al [25] reported that $T_{DL}$ was reduced by use of SDS and cetyl tri-methyl-ammonium bromide (CTAB), while use of PEG-1000 (a polymer) increased $T_{DL}$. They concluded that the occurrence of bubble jet and bubble explosion processes on superheated hydrophilic surfaces enhanced the heat transfer efficiency and delayed LFP. Similarly, Moreau et al. [26] showed that Leidenfrost droplets of aqueous solution of SDS at high temperatures undergo violent explosion. The explosion was attributed to the Plateau-Marangoni-Gibbs effect [27].

In this article, we aim to understand the effect of surfactants on the dynamic Leidenfrost phenomenon, when droplets of aqueous surfactant solutions impact on superheated substrates. We focus on understanding the intricacies of the hydrodynamics of the processes, which lead to morphed thermal aspects of the Leidenfrost behaviour in surfactant solution droplets. In our study, we explore the role of two common laboratory surfactants, SDS and CTAB. Initially, we investigated the role of surfactant concentration,

*We*, *Oh* and spreading hydrodynamics on the $T_{DL}$ during Leidenfrost boiling ($T_s \sim T_{DL}$) regime. At temperatures higher than the $T_{DL}$ ($T_s \sim 400\ ^oC$ in the present study), droplets exhibited trampoline-like behaviour and a central-jet formation during explosive boiling ($T_s > T_{DL}$), which is dependent on the impact velocity. We explore the dynamics of such bouncing and jetting behaviour. Finally, we have presented a phase map of the different boiling regimes like transition boiling ($T_s < T_{DL}$), Leidenfrost boiling ($T_s \sim T_{DL}$) and explosive boiling ($T_s > T_{DL}$) as a function of substrate temperature and *We*. We believe this study shall provide a comprehensive and mechanistic picture on the delaying of Leidenfrost effect using surfactant solution droplets.

## 2. Materials and methods

The experimental setup is similar to our earlier works on droplet impact thermo-hydrodynamics [20, 28]. A schematic diagram of the experimental setup is illustrated in fig. 1. We have used a precision control heater plate (Holmarc Opto-Mechatronics Ltd., India), with digitized temperature control to heat the substrates to temperatures ranging from 150 $^oC$ to 400 $^oC$, and to maintain near isothermal conditions on the substrate during the experiments. The test surface was a square stainless-steel block (80mm × 180mm) mounted on the heating element. A T type thermocouple probe was inserted into the stainless-steel substrate. The thermocouple measured the temperature of the substrate ~1 mm below the top surface, and was connected to the digitized temperature controller. The temperature controller is able to maintain isothermal conditions at the heating element accurate to ±2–3 $^oC$. A height adjustable, droplet dispensing mechanism (Holmarc Opto-Mechatronics Ltd., India) with digitized control was used to release the droplet from different heights, thereby allowing different impact *We* based experiments. A micro-syringe (± 0.1 μl volumetric accuracy), which is enclosed to the droplet dispenser, is used to dispense droplets of fixed volume through a flat head steel needle (22 gauge). A high-speed camera (Photron, UK) was used to capture the images of droplet impact on the hot substrate. All experiments were recorded at 4000 frames per second using a 105 mm macro lens (Nikon).

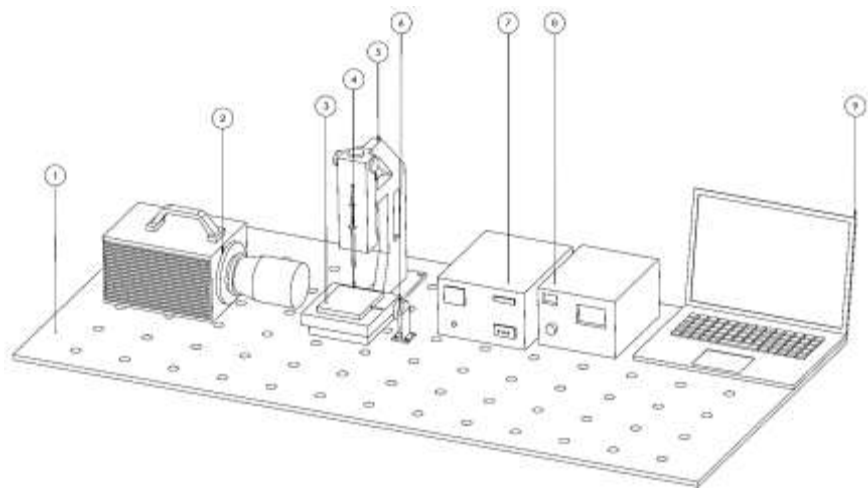

**Fig. 1:** Schematic of the experimental setup: (1) Vibration free table-top (2) high speed camera (3) hot substrate (4) micro syringe (5) droplet dispensing mechanism (DDM) unit (6)

LED backlight (7) DDM and backlight illumination controller (8) hot substrate controller (9) computer for data acquisition and camera control.

Before each experiment, the stainless-steel hot substrate was properly cleaned with DI water, followed by acetone, to remove dirt and/or contaminants from the heated surface. The substrate is heated to a particular temperature using the digitized controller, and allowed to attain steady state. At steady state, the temperature of the substrate may vary within ±2–3 °C of the set value. In this study, DI water, and aqueous solutions of SDS and CTAB (Sigma Aldrich India) were used as fluids. All surfactant solutions were prepared by continuously stirring at 600 rpm with a mechanical stirrer. In this study, we have used different concentrations of surfactant with respect to its critical micelle concentration (CMC). We have tested solutions with 0.125, 0.25, 0.5, 0.75 and 1 times the CMC concentration. The relevant thermophysical properties [12, 26] of these fluids have been tabulated in Table. 1. To characterize the impact hydrodynamics and surface interactions of the impacting droplets, we make use of certain conventional non-dimensional parameters, as shown in Table. 2.

**Table 1:** Experimental conditions and thermophysical properties of test fluid droplets: pre-impact droplet diameter $D_O$ (mm), density $\rho$ (kg/m³), surface tension $\sigma$ (mN/m), viscosity $\eta$ (mPa-s), and capillary length $\lambda_c$ (mm). In table.1 CMC represents critical micelle concentration.

| Liquid | $D_o$ | $\rho$ | $\sigma$ | $\eta$ | $\lambda_c$ |
|---|---|---|---|---|---|
| DI water | 2.64 | 995.67 | 71.03 | 0.791 | 2.69 |
| SDS (0.125 CMC –1 CMC) | 2.55 – 2 | 972.15 - 1010 | 67.725 - 39.8 | 0.792 - 0.802 | 2.66 – 2.0 |
| CTAB (0.125CMC – 1 CMC) | 2.53–2.34 | 994 - 994.25 | 49.31 - 33.53 | 0.731-0.732 | 2.24 - 1.85 |

**Table 2:** List of dimensionless parameters used: $\rho$, $V$, $D_o$, $\sigma$, $D$, $t$, $A$, $R$ and $Re$ represent density, impact velocity, pre-impact diameter, surface tension, spreading diameter, spreading time, amplitude of oscillation, radius of the impacting surfactant droplets and Reynolds number, respectively.

| Dimensionless parameter | Expression | Range of values |
|---|---|---|
| Weber number | $We = (\rho V^2 D_o)/\sigma_{drop}$ | 9.2 – 49.78 |
| Ohnesorge number | $Oh = \sqrt{We}/Re$ | 0.0015 – 0.0030 |
| Bond number | $Bo = (\rho g D_o^2)/\sigma_{drop}$ | 0.82 – 0.96 |
| spread factor | $\beta_{max} = D_{max}/D_o$ | 1.0 – 2.8 |

## 3. Results and discussions

In this section, the drop impact dynamics at high substrate temperatures will be discussed. We have varied the impact *We* by changing the release height of the droplets, the drop sizes, and the surface tension (due of different concentrations of the surfactants). The variation of drop diameters and surface tension was mentioned in table 2. The impact dynamics behavior has been classified along the lines of previous reports [24, 29-30]. The minimum temperature at which the drop levitates over a stable, thin vapour layer without splashing is considered as the Leidenfrost temperature ($T_L$) [18-24]. When the droplet is gently placed (*We* ~0) over the heated substrate, then the Leidenfrost temperature (LFT) is termed as the static Leidenfrost temperature. However, when a droplet is released from a certain height, the influence of initial impact velocity (quantified through *We*) also becomes an important factor in the Leidenfrost dynamics. Since the LFT changes with the *We*, it is common practice to term it as the dynamic Leidenfrost temperature ($T_{DL}$) [6, 13]. We have considered the $T_{DL}$ as the substrate temperature at which the impacting fluid droplet displays onset of intact rebound off the hot surface, and the corresponding boiling state is denoted as Leidenfrost boiling.

**(a) Influence of substrate temperature on spreading dynamics**

The change in surface tension (due to addition of surfactant) and the thermal condition of the substrate subsequently affects the spreading behaviour on hot surface. As mentioned in earlier studies [3, 5-6] on many applications, it is essential to understand the spreading dynamics to estimate heat transfer rate of droplets during the Leidenfrost stage. In this section, we shall discuss how surfactant concentration and substrate temperature interplay to bring about changes in the spreading hydrodynamics. Fig. 3a and 3b illustrate the maximum spread state of impacting DI water and surfactant droplets on a substrate at ambient temperature ($T_a \sim 30^oC$), and at their respective dynamic Leidenfrost temperatures ($T_{DL}$), respectively. In this context it must be noted that due to variation in equilibrium surface tension, the droplet diameter size varies with the surfactant concentration (droplet size impacting the surface decreases with increase in surfactant concentration).

As seen from table 1, the highest concentration (1 CMC SDS) has droplet diameter ~2 mm, whereas the lowest concentration (0.125 CMC SDS) solution has droplet diameter ~2.5 mm. At 0 CMC (i.e. water) the droplet diameter is ~ 2.7–2.8 mm. Since the initial drop diameters are different, it is cumbersome to quantify the effect of surfactants on the spreading diameter directly from fig. 2. However, the effect of impact velocity for a fixed surfactant concentration is readily evident from fig. 2. For a fixed surfactant concentration, maximum spreading diameter increases with increase in impact velocity at both ambient conditions and at the corresponding Leidenfrost state. It is noted from fig. 2b that compared to water at its Leidenfrost state, the tendency of crown formation and fragmentation of secondary droplets decreases with the addition of surfactants. We further note that the vigorousness of boiling during impact at the corresponding Leidenfrost state is reduced with surfactants, as evident from the reduction in the splashing behaviour at impact (fig. 2b, 1$^{st}$ column).

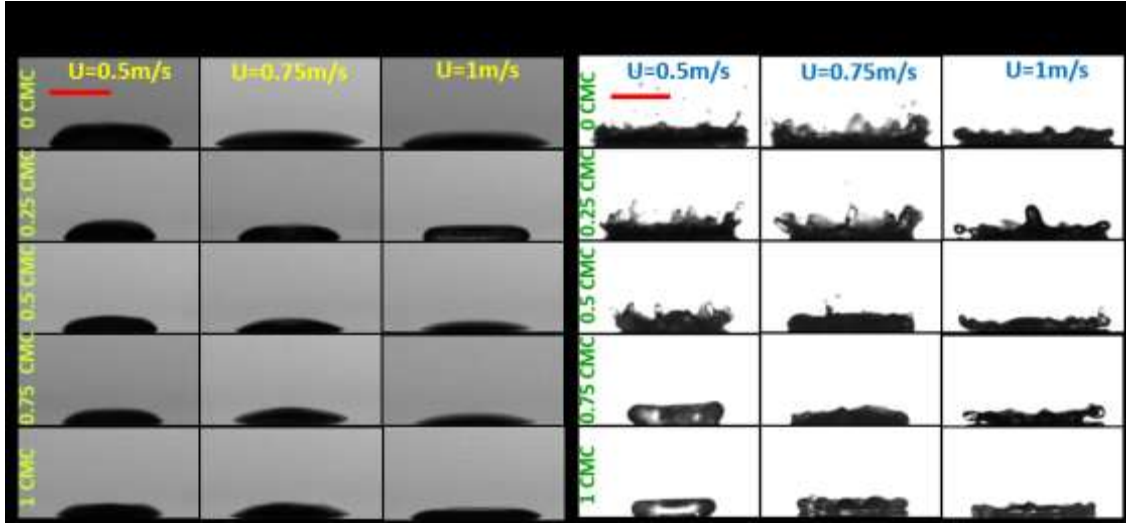

**Fig. 2:** Maximum spread state of SDS droplets for different impact velocities at (a) $T_a \sim 30\ °C$, and (b) at their respective $T_{DL}$. The scale bar represents ~2.64 mm. The measurements of the maximum spreading diameter ($D_{max}$) with varying Weber number and surfactant concentration are shown in Supplementary information (fig. SI-2). The transient dynamics of SDS droplets at 220° C (temperature at which a water droplet attains Leidenfrost state at 0.5 m/s in the present study) is provided in fig. SI-1.

**(b) Influence of *We* and surfactant concentration on spreading dynamics during Leidenfrost boiling ($T_{DL}$)**

As both maximum spreading diameter ($D_{max}$) and initial droplet diameter ($D_o$) changes with increasing surfactant concentration, we have quantified the spreading dynamics by defining the non-dimensional maximum spread factor ($\beta_{max}=D_{max}/D_o$). This section highlights the role of *We* on the spreading dynamics at the Leidenfrost state of each fluid i.e. at $T_{DL}$. The *We* based on impact velocity at the moment of impact is governed by the droplet release height, droplet diameter, as well as the surface tension. For the sake of simplicity, we have considered the equilibrium surface tension at ambient temperature [24] for calculation of the *We*. Fig. 3(a) and (b) show the variation of $\beta_{max}$ for different concentrations (0 to 1 CMC) of both SDS and CTAB solution droplets at their respective $T_{DL}$ over different *We*. For a given test fluid, the *We* is increased by increasing the drop release height. Consequently, the $\beta_{max}$ (for a particular fluid) also increases due to the dominance of the stored kinetic energy of impacting fluid droplet against the surface energy.

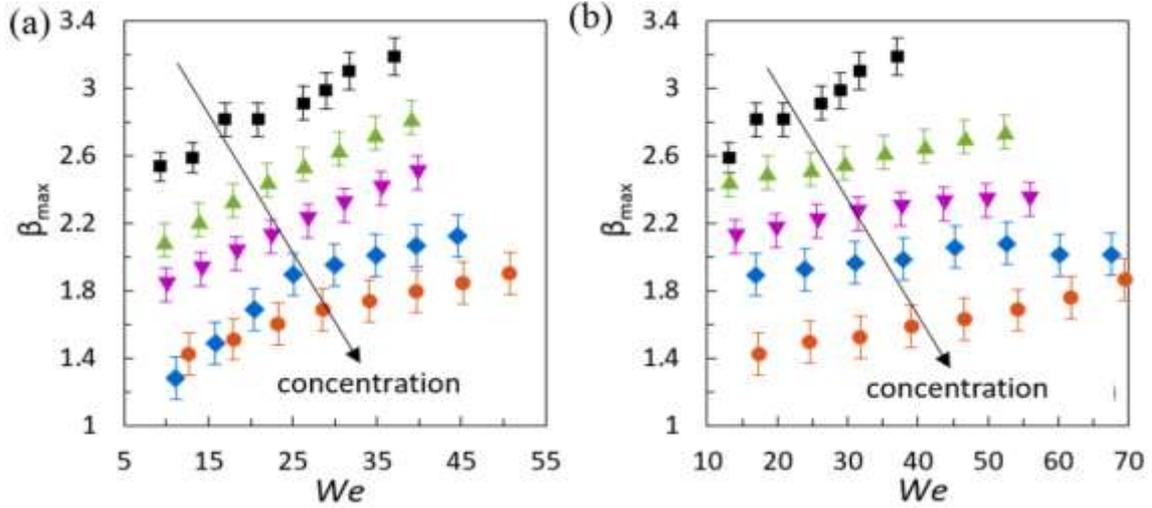

**Fig. 3.** The maximum spread factor ($\beta_{max}=D_{max}/D_o$) at different *We* for surfactant solution droplets of (a) SDS (b) CTAB, respectively. In above figures, (■) represents 0 CMC or water, (▲), (▼), (♦) and (●) represent surfactant concentrations of 0.25 CMC, 0.5 CMC, 0.75 CMC and 1 CMC, respectively. The arrows represent increasing surfactant concentration. Fig. SI-3 (supplementary information) shows the variation of $\beta_{max}$ for different impact velocities.

From fig. 3a and 3b, it is observed that $\beta_{max}$ decreases with increase of surfactant concentration, at a particular *We*. The decreasing trend of spreading diameter with increasing surfactant concentration (which equilibrium surface tension is reduced) is counter-intuitive [20, 29-30]. Recent studies have shown the role of dynamic surface tension (DST) in case of determining maximum spreading [31-32]. These studies have put forward that during the fast spreading dynamics (occurring in the order of milliseconds), the DST should be considered instead of the equilibrium surface tension. The surfactants were categorized as "*fast*" and "*slow*" depending upon the adsorption-desorption kinetics of the surfactant molecules from the bulk of the fluid to the liquid-air interface, and resultant transient reduction in surface tension. The role of DST on maximum spreading is more prominent at higher impact velocities [33]. At higher impact velocities it was observed that maximum spreading diameter of the surfactant solution droplets were less than that of water drops. Reports have suggested that for high impact velocities and resultant smaller time scale of spreading process, the occurrence of non-uniform distribution of surfactant molecules at the air-liquid interface may give rise to Marangoni stresses, which may impede the process of droplet spreading [33].

CTAB was shown to be a "*slow*" surfactant in a study by Hoffman et al [31]. Based on these studies, we hypothesize that both the surfactants (CTAB and SDS) used in our experiments are "*slow*" to reduce the interfacial force, thereby failing to increase $D_{max}$ to the intuitive levels at their respective dynamic Leidenfrost temperatures ($T_{DL}$). In addition to the potential role of DST [31-33], the atypical decreasing trend of spreading factor with increasing surfactant concentration may also be due to thermo-capillarity driven Marangoni flows [34–37]. Due to the high temperature gradient between the bottom and top surface of the droplets at the $T_{DL}$, and the resultant surface tension gradient, a toroidal convection sets in the droplet [36]. Fluid motion starts along the spherical periphery and sequentially accelerates downward to the contact point near the center. This internal convection impedes the spreading process, as the flow tries to curl up in the opposite direction instead of advancing horizontally towards the spreading contact line. We believe that such thermo-capillarity induced internal and interfacial convection becomes stronger with increasing surfactant concentration (due to additional solutal Marangoni effect [38-39]), and results in the decreasing trend of spreading factor with increasing surfactant concentration for a fixed *We*.

**(c) Role of surfactant concentration on dynamic Leidenfrost temperature($T_{DL}$)**

The dependence of dynamic Leidenfrost temperature ($T_{DL}$) on surfactant concentration is discussed in this section. Figure 4 (a) and (b) illustrate the side view and top view of the impact dynamics of various concentrations of SDS droplets at their corresponding Leidenfrost state. From fig. 4, it can be readily observed that at a fixed velocity of U=0.5m/s, the time to attain the maximum spreading diameter is decreasing with the increase of surfactant concentration. This effect is obvious as the droplet size decreases with the increase in surfactant concentration (refer table 1). The residence time (time from moment of impact to onset of lift-off) decreases with the surfactant concentration (4$^{th}$ column of fig. 4 a, b). The Leidenfrost temperature $T_{DL}$ was denoted as the temperature at which the droplet showed rebounding off the surface for the first time [20]. At a slightly lower temperature than $T_{DL}$, the droplets were in always contact wholly or partially with the heated substrate. With increasing surfactant concentration, the droplets exhibit mode shapes characterized by undulating interface (fig. 4b, 2$^{nd}$ and 3$^{rd}$ column). Such prominently distorted interface during the spreading and boiling phenomena is akin to the Taylor-Marangoni instability discussed in literature [40]. In the present case, the presence of the surfactants augments the thermal Marangoni instability due to addition of the solutal counterpart, which leads to such prominently distorted interface of the droplet due to thermo- and soluto-capillary disturbances.

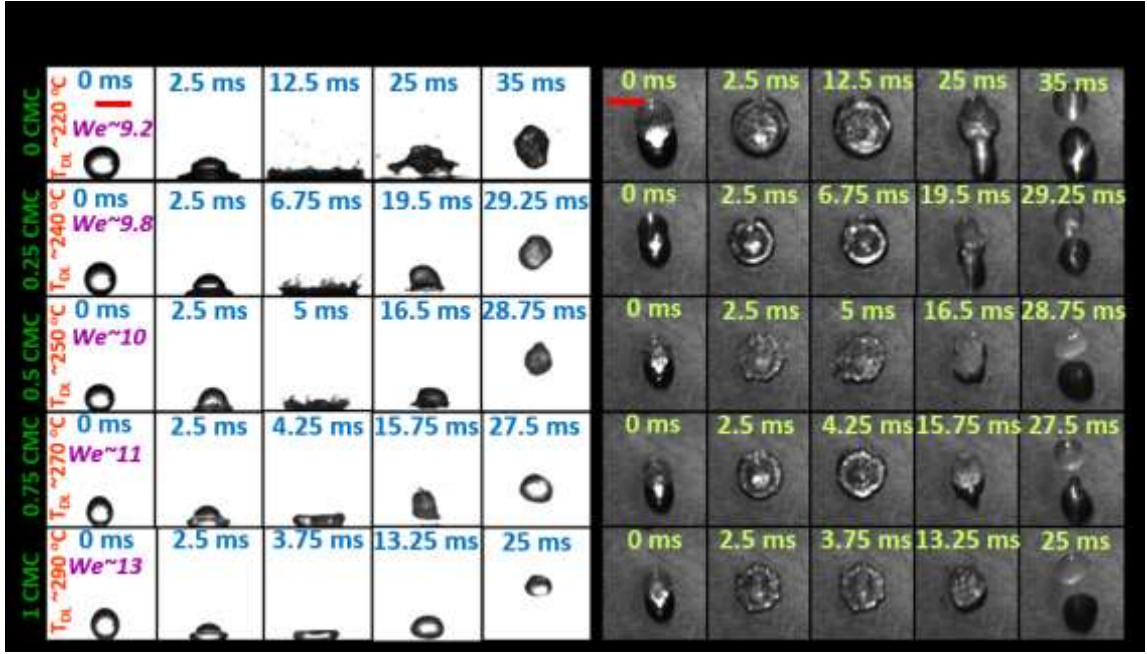

**Fig. 4.** Temporal evolution of the impacting SDS solution droplets with low impact velocity U=0.5 m/s at their respective $T_{DL}$: (a) side view (b) top view. For the same initial impact velocity, *We* is changing mildly due to variation in surfactant concentration. The scale bar represents 2.64 mm. The 3$^{rd}$ column of each row represents the instant at which maximum spread occurs. The 4$^{th}$ column of each row represents the instant at which the droplet begins to levitate for the first time.

Figure 5 a, and b show the variation of $T_{DL}$ with variation of impact velocity and surfactant concentration, for both SDS and CTAB solution droplets, respectively. For a fixed impact velocity, $T_{DL}$ increases with increase in surfactant concentration. The increasing trend is more prominent for SDS solutions (fig. 5 a). At the lowest impact velocity (0.5 m/s), the $T_{DL}$ for 1 CMC SDS solution is higher than water droplets by almost ~70 $^{o}$C, which is a high delay of the onset of Leidenfrost effect, and may have strong implications for several utilities. In case of CTAB solutions, $T_{DL}$ initially increases with increasing surfactant concentration up to 0.25 CMC, and then maintains a near-constant value at a higher temperature compared to water (0 CMC) (fig. 5b). For both the solutions, at fixed surfactant concentration, $T_{DL}$ decreases with increase in impact velocity. This trend is similar to the Leidenfrost phenomena studies with water and polymer droplets [20]. With increase in impact velocity, due to higher inertia, droplets spread to a greater distance radially (as long as the impact velocity is low enough to not induce splashing/fragmentation). The subsequent increase in area of contact between the spreading droplet and the heated surface promotes the formation of nucleation sites at liquid-solid interface to form vapor pockets. The enhanced rate of vapor formation beneath the drop helps in stable vapor layer formation and results in droplet levitation above it. Hence with increasing impact velocity, the propensity for attaining Leidenfrost state onsets at lower temperature [20]. Following this argument, the increasing trend of $T_{DL}$ with increase in surfactant concentration can be explained. Since on heated surfaces, the radial spreading is observed to decrease with the surfactant concentration, the propensity of achieving the Leidenfrost state also decreases compared to water.

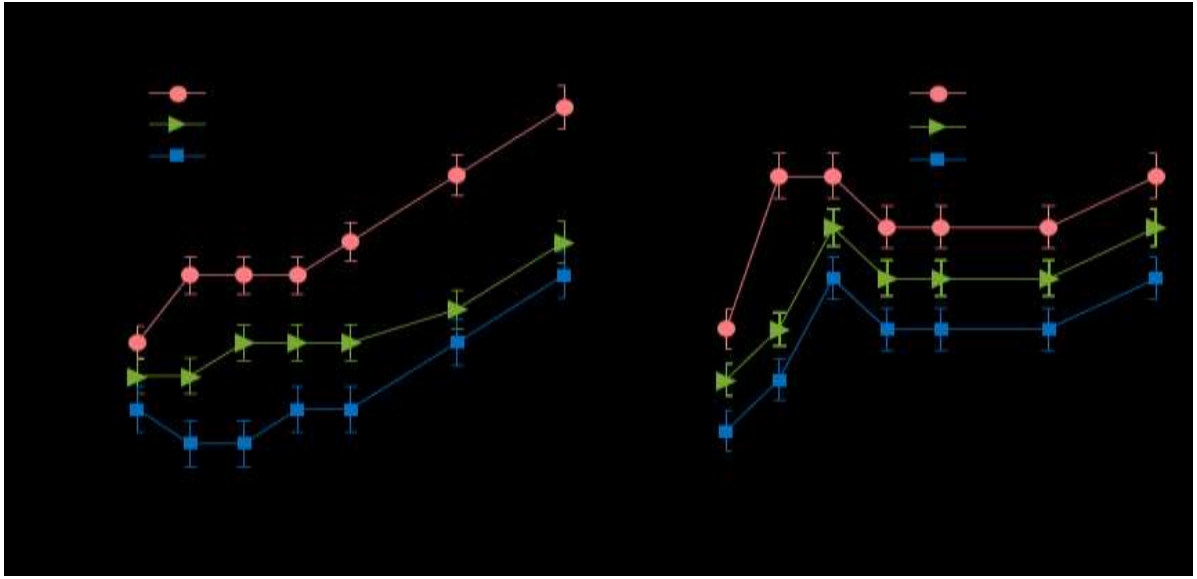

**Fig. 5:** The dynamic Leidenfrost temperature ($T_{DL}$) over surfactant concentration of (a) SDS and (b) CTAB fluid droplets.

**(d) Scaling $T_{DL}$ with Weber (*We*) and Ohnesorge (*Oh*) numbers**

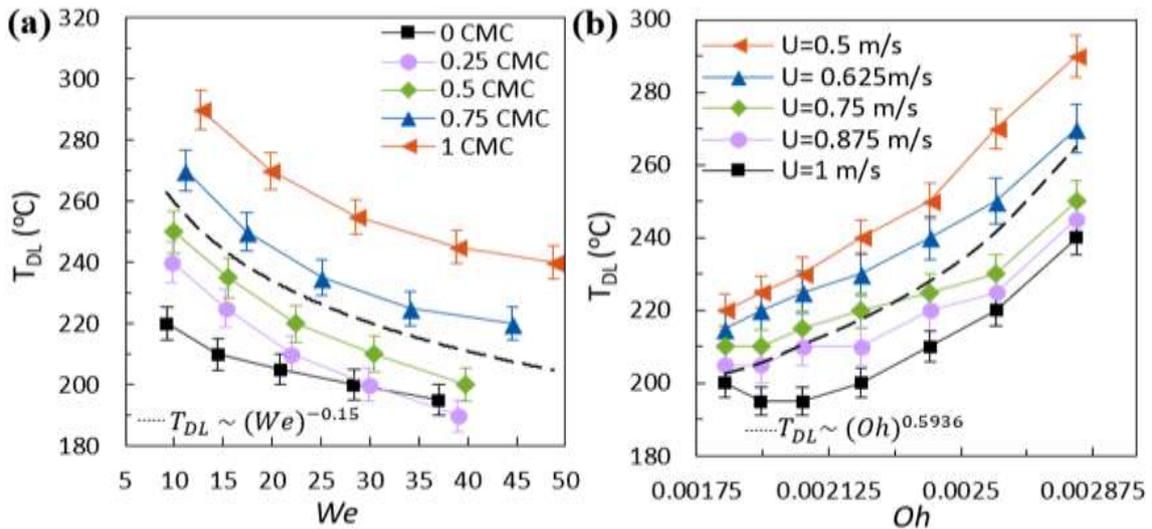

**Fig 6.:** Dynamic Leidenfrost temperature ($T_{DL}$) with respect to (a) Weber number *(We)* and (b) Ohnesorge number *(Oh)* for SDS solution droplets. The black dotted lines (-----) in(a) and (b) represents the scaling behavior of $T_{DL}$ with *We* and *Oh* respectively (as per given scaling in each figure).

Subsequently, based on the information highlighted in figure 5 a and b, the dependence of $T_{DL}$ on *We* for SDS solution droplets is shown in fig. 6(a). It is evident that the $T_{DL}$ decreases with increase of *We* for all solution concentrations. Since the $\beta_{max}$ (refer fig. 4.) increases with increase of *We*, heat transfer from the hot substrate occurs over a larger area of the thin liquid disc formed by the droplet at maximum spread state. As a result, formation of stable vapor cushion is more favourable at lower temperatures, resulting in decreasing trend of $T_{DL}$ with increasing $We$. At the same $We$, the $T_{DL}$ increases with increase of surfactant concentration. This reveals that $We$ alone is not enough to explain the effect of surfactant concentration on the $T_{DL}$. To improve upon this observation, we introduce another non-dimensional number, the Ohnesorge number ($Oh = \frac{\mu}{\sqrt{(\rho\sigma D)}} = \frac{\sqrt{We}}{Re}$). $Oh$ is the ratio of viscous forces to both surface tension and inertial forces. Fig. 6 (b) clearly reveals that $T_{DL}$ vs. $Oh$ reflects the increasing trend of $T_{DL}$ with surfactant concentration at a fixed impact velocity (fig. 7) as in both cases the equilibrium surface tension decreases with increment of $Oh$ and surfactant concentration. From figure 6, it is visually evident that $T_{DL}$ is dependent on both $We$ and $Oh$. Analogous to a previous study of Leidenfrost effect with high-alcohol surfactants (HAS) [24], we have attempted to introduce a scaling relationship of $T_{DL}$ with $We$ and $Oh$. The $T_{DL}$ converges to static Leidenfrost temperature ($T_{SL}$), when $We$ approaches zero. The experimental results of $T_{DL}$ are fit to conform to the following relationship: $T_{DL} = (aOh^b + c)We^{-d} + e$. Through least squares regression, the scaling correlation of $T_{DL}$ in terms of $We$ and $Oh$ (with $\pm 15$ % confidence interval, refer fig. 7) is obtained as: $T_{DL} = (365Oh^{0.594} + 128)We^{-0.15} + 142$. We believe that this scaling correlation will be of valuable reference for future researchers in this field.

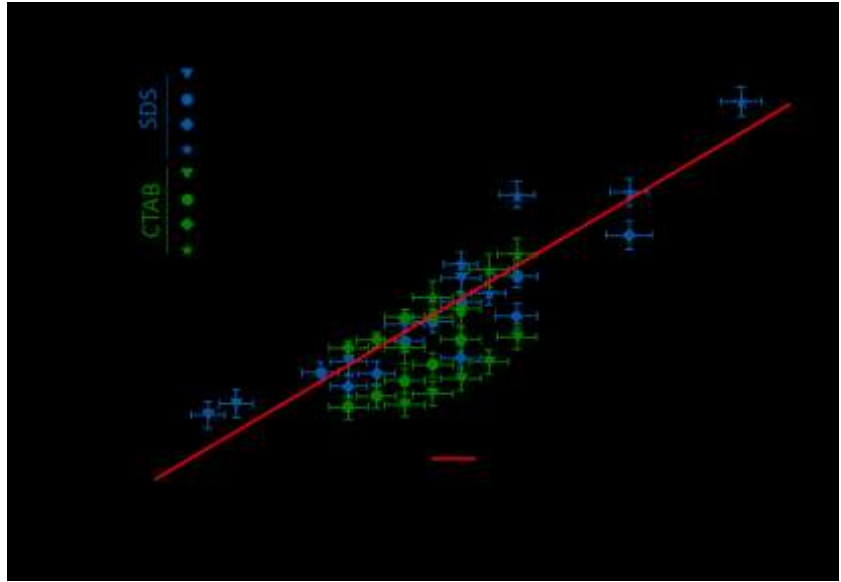

**Fig 7:** Comparison between the calculated and experimental $T_{DL}$ of impacting surfactant solution droplets with different concentrations.

### (e) Influence of Bond number (*Bo*) on $T_{DL}$

The variation of surfactant concentration significantly reduces changes the pre-impact droplet size (ref. Table 1) due to reduction in equilibrium surface tension. So, it is important to discuss the influence of Bond number ($Bo = \frac{\rho g D^2}{\sigma}$) on the $T_{DL}$. The dependence of $Bo$ during the levitation state at $T_{DL}$ on varying concentrations at a fixed impact velocity U=0.5m/s is presented in fig.8. In figure 5, it was already observed that the $T_{DL}$ increases with the surfactant concentration. Due to reduced droplet diameters for higher surfactant concentration, $Bo$ is decreasing with increasing surfactant concentration, in spite of reduction in equilibrium surface tension. As a result, figure 8 highlights that $T_{DL}$ decreases with increasing *Bo*. This figure may provide relevant information on the dynamics of the Leidenfrost droplets only in terms of the droplet and its physical properties.

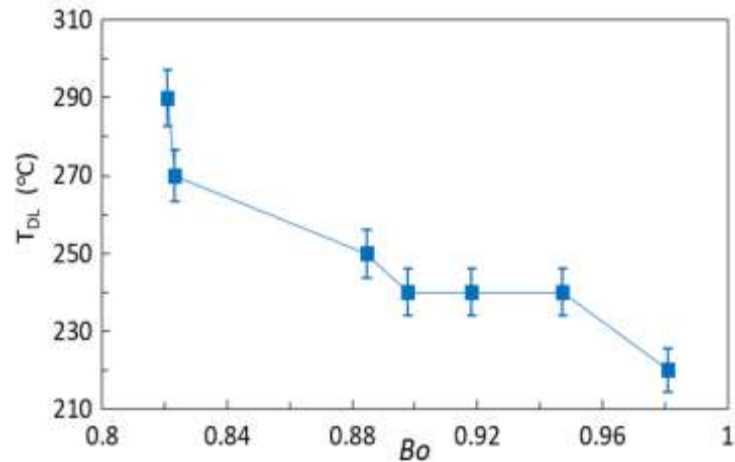

**Fig. 8.** Effect of Bond number (*Bo*) on dynamic Leidenfrost temperature $T_{DL}$ of different concentrations of SDS solution droplet at impact velocity U=0.5m/s.

### (f) Trampolining dynamics of droplet at $(T_s>T_{DL})$ ~400⁰ C

At temperatures significantly higher than $T_{DL}$ (*i.e., at* $T_s$~ 400 °C in the present discussion), droplet impact behaviour is significantly changed with impact velocity. At lower impact velocity ~0.5m/s, droplets display a series of bouncing off dynamics on the substrate. This type of behaviour is termed as trampolining dynamics, as described in literature [9, 11, 20, 40]. On further increase of impact velocity to 1m/s, a central jet formation and ejection was observed during the retraction phase. In this section we will discuss the trampolining behaviour. The time dependent bouncing behaviour of the droplets at different SDS concentrations is presented in fig. 9. At significantly high temperature ~400°C, prompt formation of the stable vapour layer (just after ~10 ms, refer 3$^{rd}$ and 4$^{th}$ column of fig. 9) enhances the cushioning effect, and the rapid vapour generation leads to the ejection of the droplet from the surface in nearly intact manner. The sustained oscillations are possible due to the thermo-capillarity induced flow and vapour generation during the contact, energizing the flow [36-40]. This is significant deviation from the impact dynamics at their corresponding $T_{DL}$ (refer fig. SI-4). At $T_{DL}$, for all impact velocities, droplets rebounded off the surface. For a fixed test fluid, rebound height decreased with increasing impact velocity. This is consistent with a previous study where the increase of viscous dissipation and resultant reduction in retraction kinetic energy with increasing impact velocity was highlighted [41].

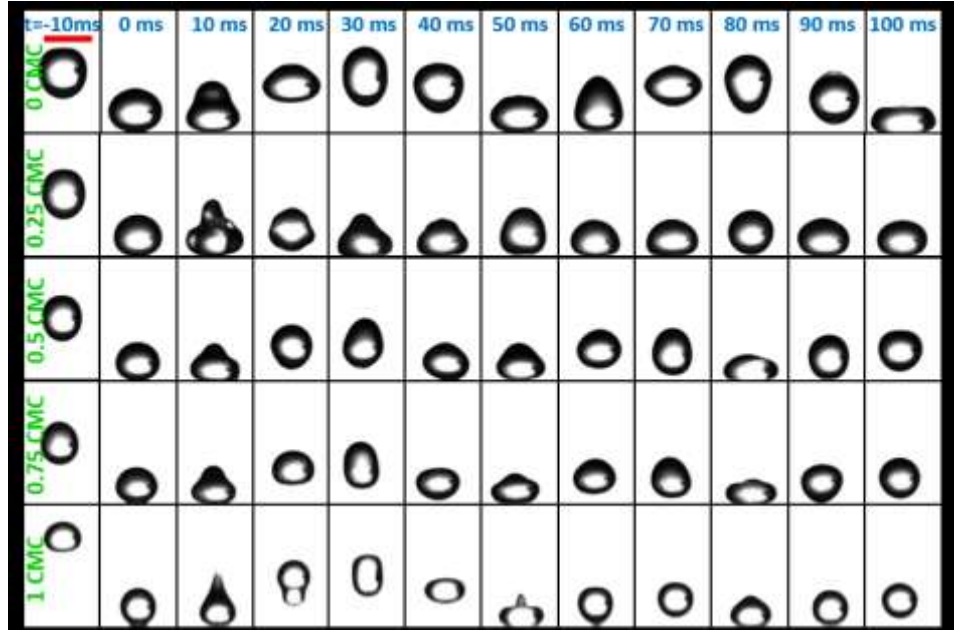

**Fig 9:** Droplet trampolining dynamics of SDS solution droplets with low impact velocity U=0.5m/s at surface temperature of 400 °C. The scale bar represents 2.64mm. The dynamics of 1 CMC SDS droplets at 0.5m/s , 0.75 m/s and 1 m/s are presented in the supplementary section in movie format.

Subsequently, the non-dimensional rebound height (rebound height/ drop radius) for the 1$^{st}$ rebound and the coefficient of restitution (rebound height /initial height)$^{0.5}$ for 0.5 and 0.75 m/s is presented in figure 10 a and b, respectively. Similar to the rebound height at $T_{DL}$ s (SI-4), rebound height reduces with increasing impact velocity for 0.5 m/s. However, the rebound height measured for higher velocity (0.75 m/s) shows monotonically decreasing trend with increasing surfactant concentration. Also, the non-dimensional rebound height is higher for 0.75 m/s velocity unlike the trend presented in SI-4 at respective $T_{DL}$ s. Compared to the 0.5 m/s case, the frequency of rebound is reduced for 0.75 m/s (refer supplementary movies). The coefficient of restitution is defined as $\varepsilon = \sqrt{\frac{h_{rebound}}{h_{impact}}}$. Fig. 10b presents the coefficent of restitution with variation in imapct velcoity and varying SDS concentrations. The ε is much less compared to the previous studies on Leidenfrost trampolining behaviour where the ε was almost equal to 1 [41]. At lower impact velocity, the coefficient is noted to decrease with concentration, and then increase, and this behavior is observed in repeated experiments. At lower impact velocity, the low kinetic energy allows for better heat transfer and vapour generation, which in turn promotes the Taylor-Marangoni instability at the droplet interface (refer fig. 4b) [40]. We believe that onset of this instability reduces the available kinetic energy during cushioning and rebound, leading to reduced coefficient of restitution at lower concentrations. As the surfactant concentration increases, the effective spreading reduces. Consequently, the available interfacial area for the instability to manifest reduces, thereby reducing its strength. We believe this is the reason why the droplet is further able to exhibit increased coeffieicnt of restitution.

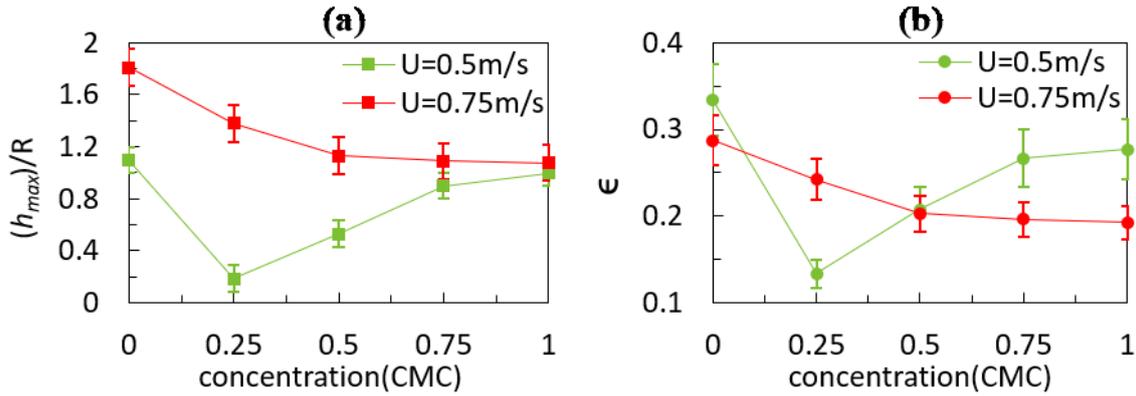

**Fig. 10 (a)** Non-dimensional rebound height vs. SDS concentrations for different impact velocities. Trampolining dynamics over time is presented as SI-5 **(b)** coefficient of restitution, $\mathcal{E}$, for the same process in (a).

**(g) Fragmentation and jetting dynamics at ($T_s > T_{DL}$) ~400°C**

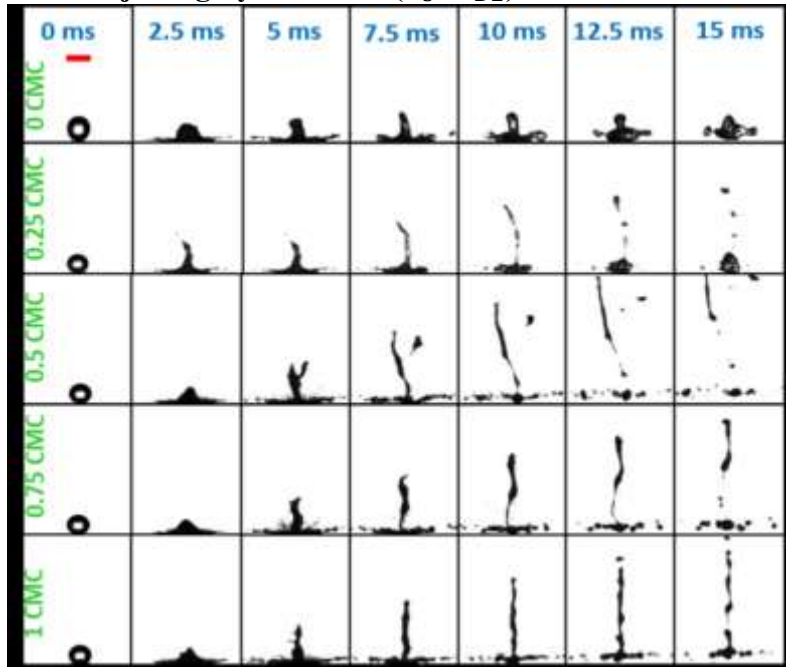

**Fig. 11.** Jet ejection dynamics of SDS solution droplets at high impact velocity U=1 m/s for surface temperature ~400 °C. The scale bar represents 2.64 mm.

At higher impact velocity ~1m/s, and very high surface temperature, the droplets no longer stay intact and fragment into secondary droplets during the recoiling process after maximum spreading. This impact outcome is termed as explosive boiling. During retraction, the droplets undergo fragmentation into smaller droplets. Formation of a central jet was noticed and the height of the jet was observed to increase with the increase in SDS concentration (refer fig. 11). It is noteworthy that due to higher impact velocity, the process of jet formation has started just after ~5 ms from the instant of the droplet touching the substrate, even before the attainment of maximum spread state. A previous study by Siddique et al [42] showed that the jet formation was due to the inertial collapse of the air cavity formed beneath the droplet, caused by the rapid vaporization at the point of contact due to very high surface temperature. From fig. 11 and fig. 12 a, it is evident that the central jet height increases with increase of SDS concentration. For further description of jet ejection dynamics, the role of critical substrate temperature ($T_s$) and $We$ at which the jet formation was observed for first time has been highlighted in fig.12 b and 12 c, respectively. Fig. 12 (b) reveals that the critical $T_s$ increases with increase of surfactant concentration. Fig. 12(c) shows that the critical $We$ increases with increase of surfactant (SDS) concentration. The higher surfactant concentration droplet has less surface tension than low surfactant concentration droplet. When a droplet has low surface tension, it needs more stored kinetic energy to propagate the jet during the retraction phase. Similarly, the lesser stored kinetic energy is quite sufficient to support the onset of central-jet formation. Thus, higher surfactant (SDS) concentration droplets need higher critical $T_s$ and large critical $We$ to exhibit the onset of central jet-formation.

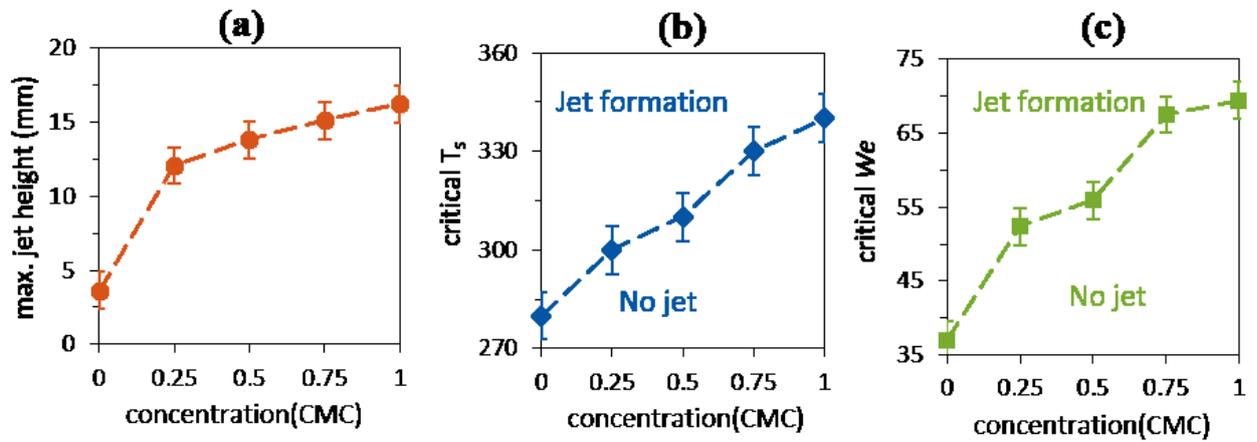

**Fig. 12.** Jet formation dynamics: (a) maximum jet height (b) critical substrate temperature ($T_s$) and (c) critical $We$ with respect to SDS concentration of impinging droplets with high impact velocity U=1m/s at surface temperature 400 °C.

**(h) Boiling regimes of impacting surfactant droplets**

Finally, we have prepared phase maps of different boiling regimes of the impacting water, and surfactant solution droplets of SDS and CTAB on hot substrate in fig. 13 (a), (b) and (c) respectively, along the lines of previous studies [1, 15]. The parametric dependence of substrate temperature (150°C-400°C) and Weber number based on the impact velocities (0.5 to 1 m/s) on the various boiling regimes is presented in the fig. 13. The boiling regimes were distinguished into transition boiling, Leidenfrost state, trampolining behaviour and explosive boiling, marked with colored letters I (blue), II (red), III (pink) and IV (green) respectively in fig. 13. In comparison to DI water droplets (see fig.13 (a)), the phase diagram of SDS and CTAB droplets boiling behavior (presented in fig.13 b & c respectively) highlights the greater coverage of transition boiling regime due to increment of the $T_{DL}$ with addition of surfactants. Also, the upper temperature limit of transition boiling reduces with $We$ as $T_{DL}$ was observed to decrease with $We$. Similarly, the explosive boiling region with central jet ejection is also occurring over a wider phase space, due to higher chances of jet formation and fragmentation in the surfactant solutions. In this context, it is worthwhile to discuss the competition between both transition boiling and explosive boiling regimes against impact *We*. On one hand, we notice that both SDS and CTAB solution droplets shows onset of transition boiling at lower substrate temperature even at higher *We*. On the other hand, explosive boiling is exhibit at higher substrate temperatures even at low *We*. Thus, finally, from fig.13., the overall Leidenfrost regime is suppressed with increase of surfactant concentration. For a certain fixed surfactant concentration, the Leidenfrost regime also shrinks (reduced shaded area) further with increase in impact *We*.

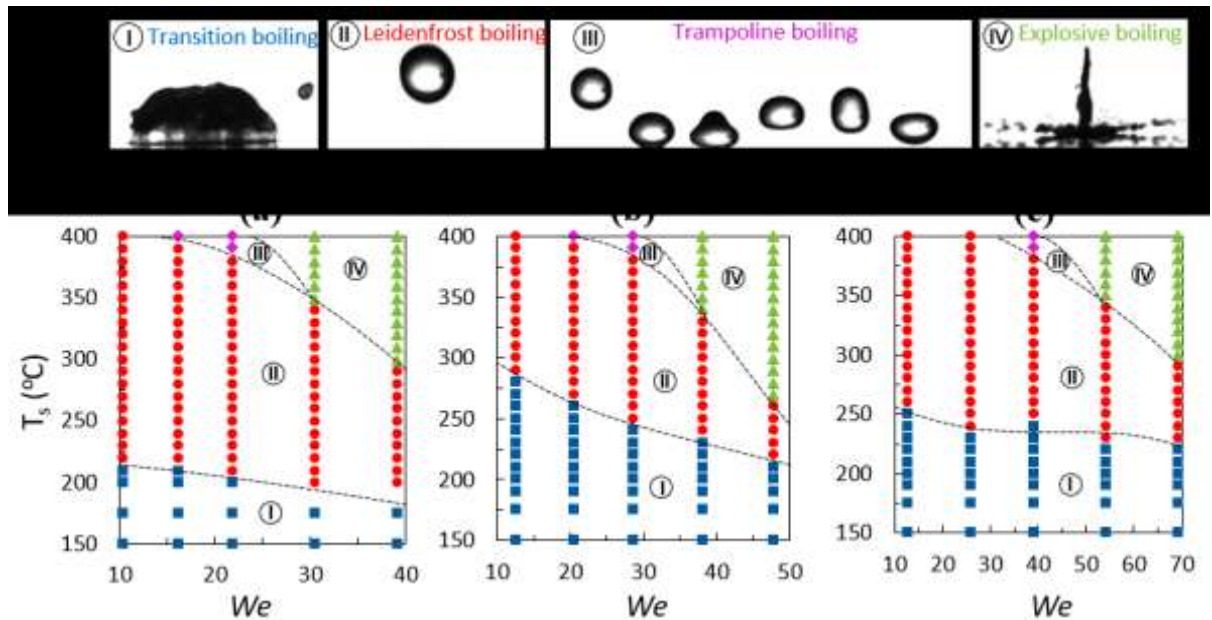

**Fig. 13.** Different boiling regimes impacting droplets of (a) water (b) SDS, and (c) CTAB. The top row shows representative snapshots of the different boiling behaviors.

## 4. Conclusions

The present study showed that both the surfactants (SDS and CTAB) are effective in increasing the Leidenfrost point $T_{DL}$ compared to water. The $T_{DL}$ was decreasing with Weber number for a fixed concentration, whereas, the $T_{DL}$ was increasing with surfactant concentration. Scaling relationships of $T_{DL}$ with *We* and *Oh* was proposed. At temperature significantly higher than $T_{DL}$ (400° C), the drop impact behavior was shown to be dependent on the impact velocity. At lower impact velocity ~0.5 m/s, the droplets were exhibiting trampolining dynamics. On increasing the impact velocity ~1m/s, droplets fragmented into secondary droplets along with the emergence of a vertical jet in the central region during the recoiling process. Finally, we have prepared a phase map of the different boiling states like transition, Leidenfrost, trampolining dynamics and explosive boiling (with fragmentation and central jet formation) as function of temperature and *We*. We believe our study will be helpful and instrumental in carrying out experiments and numerical simulations of surfactant droplet dynamics on heated substrates with further details. Also, the findings may have strong implications in design and development of safer and more reliable high temperature thermal management strategies in certain niche utilities.

**Conflicts of interests:** The authors do not have any conflicts of interest with any individuals or agencies with respect to the current research work.

**Data availability:** All data pertaining to this research work are available in this article and the supporting information document.

**Supporting information:** This document contains additional data on the hydrodynamics and behavior of the droplets during Leidenfrost effect.

**Acknowledgments:** GVVSVP would like to thank the Ministry of Education, Govt. of India, for the doctoral scholarship. DS would like to thank IIT Ropar for partially funding the work (vide grant 9-246/2016/IITRPR/144). PD thanks IIT Kharagpur (vide grant SFI) and Science and Engineering Research Board (SERB) (vide grant SRG/2020/000004) for partially funding the work.

# Supplementary Information

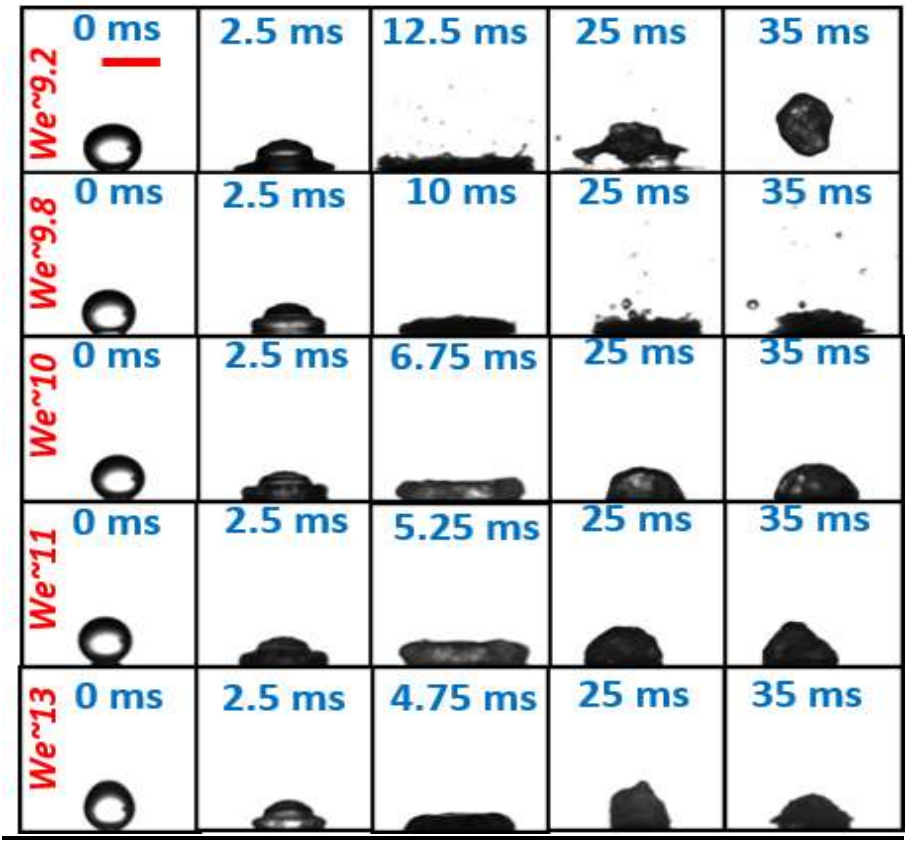

**Fig. SI-1:** Temporal dynamics of water, 0.25, 0.5, 0.75 and 1 CMC SDS solution droplets (from top to bottom) at 220° C for an impact velocity of 0.4 m/s. The 3rd column of each row represents the maximum spreading state.

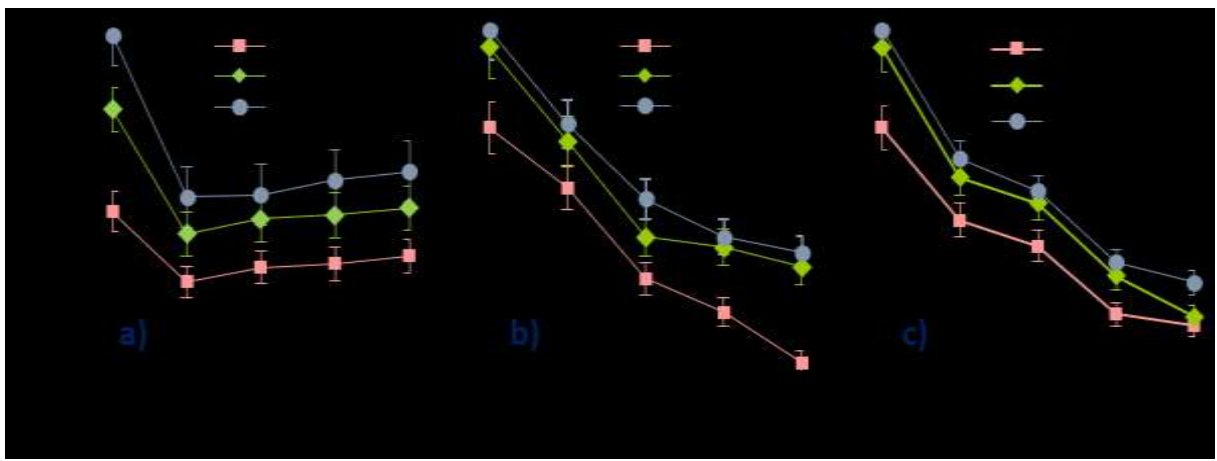

**Fig. SI-2:** Variation in maximum spreading diameter ($D_{max}$) for different impact velocities of (a) SDS solution droplets at ambient temperature ($T_a \sim 30\ ^oC$), (b) SDS solution droplets and (c) CTAB solution droplets at their respective $T_{DL}$.

At ambient temperature, the maximum spreading diameter $D_{max}$ is lesser in case of the surfactant solutions than water droplets. After the initial decrease of $D_{max}$ from water (0 CMC) to 0.25 CMC, increase in the surfactant concentration has negligible effect on $D_{max}$ at Ta~30° C. It is also evident that for a fixed surfactant concentration, $D_{max}$ increases with impact velocity. Contrary to common intuition of surfactant addition causing an increase in maximum spreading diameter due to reduction in surface tension, our experimental results showed lesser $D_{max}$ in case of surfactants laden droplets compared to water droplets at their respective $T_{DL}$s for both the surfactants (fig. SI-1b and c).

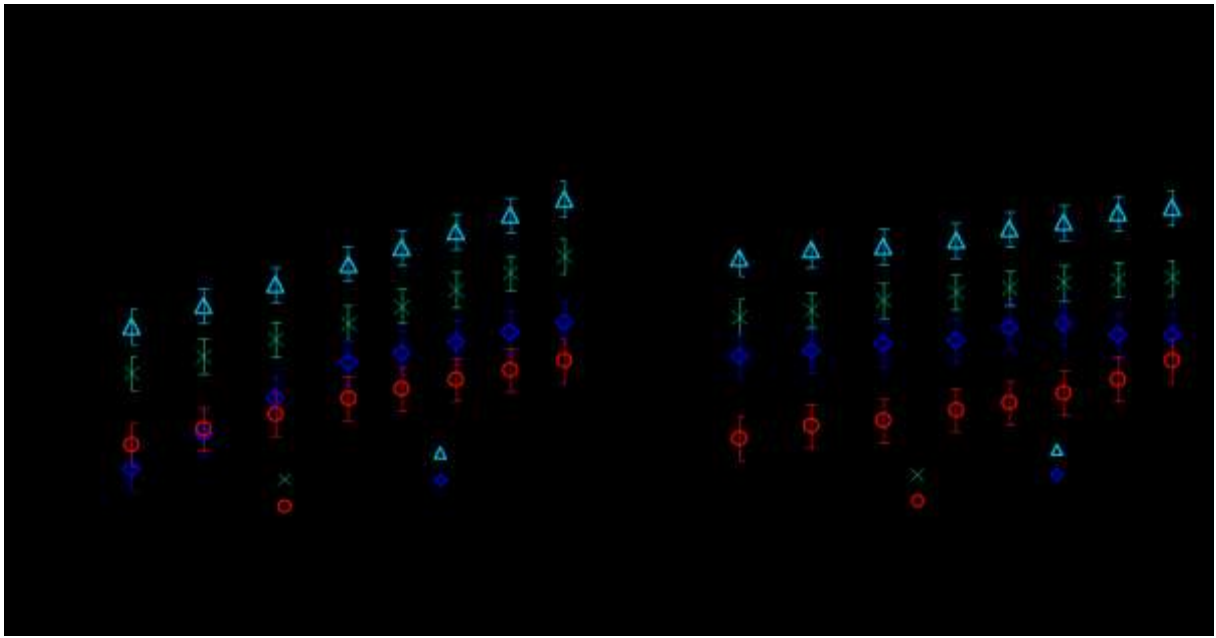

**Fig. SI-3:** Variation in non-dimensional maximum spreading diameter ($\beta = D_{max}/D_o$) over different impact velocities and different surfactant concentration of (a) SDS and (b) CTAB surfactant fluid droplets at their respective $T_{DL}$'s

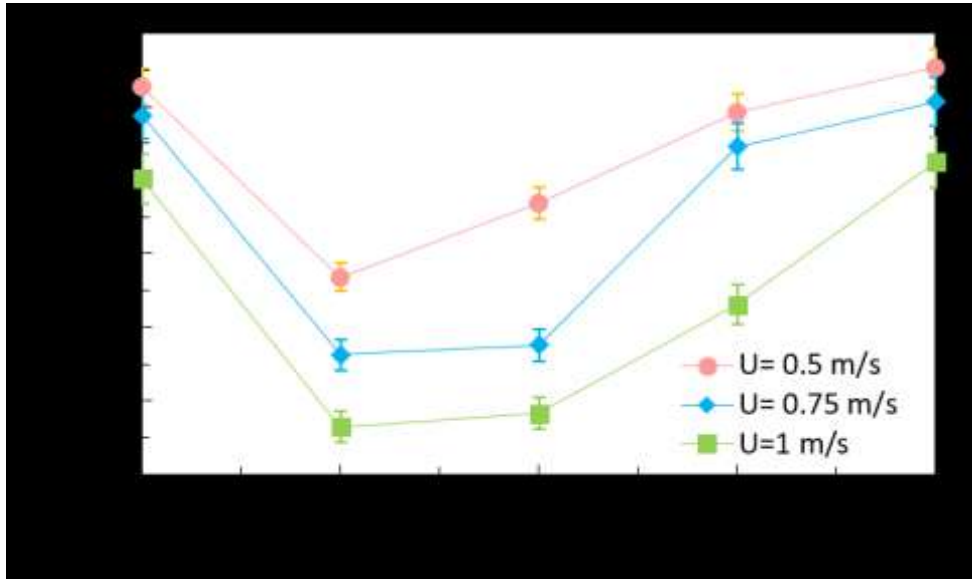

**Fig. SI-4:** Impact height of first rebound of various surfactants at their corresponding $T_{DL}$.